\documentclass[twocolumn, 11pt, reprint]{revtex4-1}
\usepackage{graphicx}
\usepackage{color}

\begin{document}

\title{ Constraining decaying dark matter with neutron stars}

\author{M. \'Angeles P\'erez-Garc\'ia$^1$~\footnote{mperezga@usal.es} and Joseph Silk$^{2, 3, 4}$~\footnote{silk@iap.fr}}

\affiliation{$^1$ Department of Fundamental Physics, University of Salamanca, Plaza de la Merced s/n 37008 Spain\\ $^2$Institut d'Astrophysique,  UMR 7095 CNRS, Universit\'e Pierre et Marie Curie, 98bis Blvd Arago, 75014 Paris, France\\ $^3$Department of Physics and Astronomy, The Johns Hopkins University, Homewood Campus, Baltimore MD 21218, USA\\
$^4$Beecroft Institute of Particle Astrophysics and Cosmology, Department of Physics, University of Oxford, Oxford OX1 3RH, UK,\\
}

\date{\today}

\begin{abstract}
The amount of decaying dark matter, accumulated in the central regions in neutron stars together with the energy deposition rate from decays, may set a limit on the neutron star survival rate against transitions to more compact objects provided nuclear matter is not the ultimate stable state of matter and that dark matter indeed is unstable. More generally, this limit sets constraints on the dark matter particle decay time, $\tau_{\chi}$.  We find that in the range of uncertainties intrinsic to such a  scenario, masses $(m_{\chi}/ \rm TeV) \gtrsim 9 \times 10^{-4}$  or  $(m_{\chi}/ \rm TeV) \gtrsim 5 \times 10^{-2}$ and lifetimes ${\tau_{\chi}}\lesssim 10^{55}$ s and ${\tau_{\chi}}\lesssim 10^{53}$ s can be excluded in the  bosonic or fermionic decay cases, respectively, in an optimistic estimate, while more conservatively, it decreases $\tau_{\chi}$ by a factor $\gtrsim10^{20}$. We discuss the validity under which these results may improve with other current constraints.
\end{abstract}

\maketitle

Disentangling the nature of dark matter (DM) is one of the greatest current challenges in physics. Whether this is realized  through a stable or a decaying particle remains unknown to date. There is a vast literature with many well-motivated particle physics models containing unstable, long-lived DM  particle candidates, see e.g.\cite{review} for a review.  Possible DM decay time-scales, $\tau_{\chi}$, are constrained by  cosmic microwave background (CMB) anisotropies, Fermi LAT limits on galaxy clusters and galactic $\gamma$-ray diffuse emission, antiprotons and the observed excess in the cosmic electron/positron flux \cite{dugger}\cite{ibarra13}. It is usually assumed that the decay daughter particles are (nearly) massless although a more generic situation with arbitrary non-zero masses, $m_D$, may also occur \cite{aoyama, salcedo, peter}.
The spread of the current bounds on the DM lifetime $\tau_{\chi}$ or, equivalently, on the DM decay rate $\Gamma_{\chi}=1/\tau_{\chi}$ is large. For example, PAMELA \cite{pamela} and Fermi LAT \cite{fermi} data can be interpreted in a scenario where a decaying $\chi$-particle has a lifetime $\tau_{e^+e^-}\sim10^{26}$ s  for DM masses $m_{\chi}\gtrsim 300$ GeV and well into the TeV range \cite{ibarra09} (we use $c=1$). Such lifetimes may appear in the context of supersymmetric grand unification theories through dimension 6 operators \cite{nardi} with  $\tau_{\rm GUT} \sim 10^{27}\, s \left(\frac{\rm TeV}{m_{\chi}}\right)^5 \left(\frac{ M_{\rm GUT}}{2\times 10^{16} \, \rm GeV}\right)^4$. On the other hand, CMB data provide a constraint $\Gamma_{\chi}^{-1}\gtrsim 30\, \rm Gyr$ for massless  daughter particles while for sufficiently heavy ones, $m_D\lesssim m_{\chi}$, decay times remain unrestricted \cite{aoyama}.

Here we consider a scenario where weakly interacting scalar bosonic or fermionic metastable DM is gravitationally accreted onto neutron stars (NSs), and possibly first onto the  progenitor stars. In brief, NSs are astrophysical objects believed to have a central core, which constitutes the bulk of the star and where mass densities are supranuclear, i.e. in excess of $\rho_0\simeq 2.4 \times10^{14}\, \rm g/cm^3$. Although there is a rich phenomenology on the possible internal core composition, for the sake of simplicity, we conservatively consider it here to be  composed of nucleon (n) fluid, with mass densities  $\rho_n\sim (1-10) \rho_0$. Under these conditions, NSs are efficient DM accretors. They can effectively capture an incoming weakly interacting $\chi$-particle passing through the star since its mean free path is much smaller than the typical NS radius. Explicitly, $\lambda_{\chi}\simeq \frac{1}{\sigma_{\chi n} n_n}$ where $\sigma_{\chi n}$ is the $\chi-$nucleon elastic scattering cross-section that we will use in the S-wave approximation and $n_n=\rho_n/m_n$ is the nucleon particle density, with $m_n$ the nucleon mass. 

Compilation of the latest results in direct detection searches \cite{snowmass} allows analysis to set limits at a level of $\sigma_{\chi n} \simeq 10^{-44}\,\rm cm^2$ in the $m_{\chi}\sim(10-10^4)$ GeV range. For the sake of discussion in this work, we will consider this $\sigma_{\chi n}$ value as representative for a DM particle candidate having in mind that current experimental efforts can potentially provide more stringent lower values.
In the same fashion, assuming an average NS with typical radius $R\simeq 12$ km and mass $M\simeq 1.5 M_{\odot}$ each DM particle will scatter inside a number of times given by
\begin{equation}
R/\lambda_{\chi} \simeq 6.1 \left(\frac{R}{12\,\rm km}\right) \left(\frac{\sigma_{\chi n}}{10^{-44} \,\rm cm^2}\right) \left(\frac{\rho_n} {3 \rho_0} \right).
\end{equation}

However, accretion of DM will proceed not only during the NS lifetime, but also in the previous late stages of the progenitor star where the dense nuclear ash central core allows the build-up of a $\chi$-distribution, $n_\chi(r)$, over time. In previous works we have considered the effect of a self-annihilating DM particle on the internal NS dynamics \cite{perez10,perez12, perez13, kumiko-perez-silk} but here we will focus on the possibility that the only process depleting DM is decay.

The picture is thus similar to proton decay searches such as those performed by Super Kamiokande \cite{sk} consisting of a 50 kton water detector looking for proton decay channels $p\rightarrow e^+\pi_0, {\bar \nu}K^+$. This volume involves $\sim 10^{33}$ protons and is thus sensitive to $\tau_p\sim 10^{33-34}$ yr as showed by analysis of data \cite{sk-decay}. By analogy, we consider the NS core as a tank of  DM that builds up in time starting as early as the progenitor star phase. The numbers of captured particles will depend on the size, mass and composition  of the progenitor star (as well as on the NS phase) once the DM-ordinary matter scattering cross-section is fixed.  

The DM accretion process onto NSs has been previously estimated, see for example \cite{gould, lavallaz}, by means of the capture rate, $C_{\chi}$, given an equation of state for regular standard-model matter in the interior of the NS at a given galactic location and with a corresponding ambient DM density. 
Taking as reference a local value for  DM density $\rho^{ambient}_{\chi,0}\simeq 0.3$ $\rm \frac{GeV}{cm^3}$, it is approximated by
\small
\begin{equation}
C_{\chi}\simeq \frac {3 \times 10^{22}}{f\left(\frac{M}{R}\right)} \left(\frac{M}{1.5 M_{\odot}}\right) \left(\frac{R}{10\,\rm km}\right) \left(\frac{1\, \rm TeV}{m_{\chi}}\right)\left(\frac{\rho^{ambient}_{\chi}}{0.3 \rm \frac{GeV}{cm^3}}\right)\,\,\rm s^{-1},
\end{equation}
\normalsize
with $f\left(\frac{M}{R}\right)={1-0.4 \left(\frac{M}{1.5 M_{\odot}}\right) \left(\frac{10 \,\rm km}{R} \right)}$ a redshift correction factor. Following \cite{kouv} we do not consider a reduction of the DM-n cross-section since we use current experimental sensitivity $\sigma_{\chi n} \simeq 10^{-44}\,\rm cm^2$, well above the geometrical cross-section. It is important to note that this dependence sets a limit on the intrinsic capability of NSs to accumulate a critical amount of DM and possibly serve as a test-bench for DM properties.

In the NS, the DM particle number, $N_{\chi}$, can be obtained solving the differential equation $\frac{dN_{\chi}}{dt}=C_{\chi}-\Gamma N_{\chi},$ considering competing processes, capture and decay, the latter treated via a generic decay rate $\Gamma$. The 
DM population at time $t$ is thus
\small
\begin{equation}
N_{\chi}(t)=\frac {C_{\chi}}{\Gamma}+ \left(N_{\chi} (t_{\rm col})- \frac {C_{\chi}}{\Gamma}\right) e^{-\Gamma (t-t_{\rm col}) }, \,\,\,t > t_{\rm col}.
\label{eqnchi}
\end{equation}
\normalsize
This solution takes into account the possibility of an existing DM distribution in the progenitor star before the time of the collapse, $t_{\rm col}$, that produced the supernova explosion.  Depending on the ${\chi}$-mass and thermodynamical conditions inside the star, it may be possible to thermally stabilize a DM internal distribution. For the $\sigma_{\chi n}, m_{\chi}$ range values discussed so far this is indeed the case.

The DM particle density takes the form $n_{\chi}(r, T)=\frac{\rho_{\chi}}{m_{\chi}}=n_{0,\,\chi} e^{-\frac{m_{\chi}}{k_B T}\Phi (r)}$, with $n_{0,\,\chi}$ the central value. $\Phi(r)=\int_0^r \frac{GM(r') dr'}{{r'}^2}$ is the gravitational potential. Assuming a constant baryonic density in the core $M(r)=\int_{0}^r \rho_n 4 \pi r'^2 dr'$, we finally obtain
\begin{equation}
n_{\chi}(r, T)=n_{0,\,\chi} e^{-(r/r_{\rm th})^2},
\label{nchi}
\end{equation}
with a thermal radius $r_{\rm th}= \sqrt{\frac{3 k_B T}{2 \pi G \rho_n m_{\chi}}}$. 

In order to asses the importance of the progenitor DM capture efficiency and thus the $N_{\chi} (t_{\rm col})$ value, let us consider a $15M_{\odot}$ progenitor star and its composition through the burning ages \cite{presn}. After the He-burning stage for $t_{\rm He \rightarrow CO}\simeq 2\times 10^6$ yr, a CO mass $\sim2.4M_{\odot}$ sits in the core with a radius $R \sim10^8$  cm. The gravitationally-captured DM population is $C^{{\rm He \rightarrow CO}}_{\chi}t_{{\rm He \rightarrow CO}} \simeq 3.35\times 10^{39} \left(\frac{1\, \rm TeV}{m_{\chi}}\right)\left(\frac{\rho^{ambient}_{\chi}}{0.3 \,\rm GeV/cm^3}\right)$. Coherence effects may play a role for slowly moving, low ${m_{\chi}}$ incoming DM particles when their associated de Broglie wavelength is comparable to the nuclear size, and in this case one should include a multiplicative factor to account for nucleus (N) instead of nucleon scatterings, i.e.  $\sigma_{\chi N}\simeq A^2 \left(\frac{\mu}{m_n}\right)^2 \sigma_{\chi n}$ where $A$ is the baryonic number and $\mu$ the reduced mass for the $\chi-N$ system. Since the later burning stages proceed rapidly, the ${\rm He \rightarrow CO}$ stage gives the main contribution to the DM capture in the progenitor. 

As the fusion reactions happen at higher densities and temperatures, the DM thermal radius contracts. In this way, for example, for $m_{\chi}=1 \,$TeV in the ${\rm He \rightarrow CO}$, $r_{\rm th}\simeq 470$ km, while  for  ${\rm Si \rightarrow FeNi}$, $r_{\rm th}\simeq 70$ km. The thermalization time $t^{-1}_{\rm th}=  \left(\frac{3k_BT}{m_{\chi}}\right)^{1/2}\sigma_{\chi N} \frac{ n_N m_{\chi} m_N} {(m_{\chi}+m_N)^2}$,  where $n_N=\frac{\rho_N}{m_N}$, for both cases is small compared to the dynamical burning timescales $t_{\rm th}/t_{\rm He \rightarrow CO}\simeq 10^{-5}$,    $t_{\rm th}/t_{\rm Si \rightarrow FeNi}\simeq 10^{-7}$. 
However during the core collapse, the dynamical timescale involved is $\Delta t_{\rm dyn\,col}\simeq \sqrt{\frac{3}{8 \pi \bar \rho G}}\simeq 10^{-3}$ s where $\bar \rho$ is an average matter density. Assuming a proto-NS (PNS) forms with $T\simeq 10$ MeV, central density $n_n=5n_0$ and a neutron-rich  fraction $Y_{\rm neut}\sim 0.9$, $n_{\rm neut}=Y_{\rm neut} 5n_0\simeq \frac{p^3_{\rm F,neut}}{3\pi^2}$, thermalization time in this phase takes longer to be achieved \cite{goldman} $t_{\rm th}=  \left(\frac {2 m^2_{\chi}}{9 m_n k_B T} \frac{p_{\rm F,neut}}{m_n} \frac{1}{n_n \sigma_{\chi n}}\right)\simeq 10^{-2}$ s.

The core collapse may thus affect the DM population inside the star since those DM particles remaining outside the PNS may effectively not be considered to play a role in the NS phase. The number of DM particles in the star interior, $r<{R_*}$, is written as $N_{\chi}=\int_0^{R_*}n_{0,\,\chi} e^{-(r/r_{\rm th})^2}dV$ and it is a dynamical quantity since $r_{\rm th}$ is temperature (time)-dependent. As long as $R_* \gg r_{\rm th}$, we obtain  $N_{\chi}=n_{0,\,\chi} (\pi r_{\rm th})^3$. The retained fraction is
\begin{equation}
f_{\chi}={N^{-1}_{\chi}}{\int_{0}^{R_{\rm PNS}} n_{0,\,\chi} e^{-(r/r_{\rm th})^2}dV},
\label{f}
\end{equation}
so that for a $R_{\rm PNS}\simeq 10$ km, $f_{\chi}\simeq 2\times 10^{-3}$. The retained DM population in the PNS after the collapse is thus $N_{\chi}= N_{\chi}(t_{\rm col}) f_{\chi}\simeq 6.7 \times 10^{36} \left(\frac {f_{\chi}}{2\times 10^{-3}}\right) \left(\frac{1\, \rm TeV}{m_{\chi}}\right)$. Let us note that the central DM density in the newly formed PNS $ n_{0,\,\chi}\simeq 3\times 10^{23}\,\rm cm^{-3}$ is much smaller than  that in the baryonic medium $\sim 10^{38}\,\rm cm^{-3}$.

At this point one should check that the DM content does not exceed the fundamental Chandrasekar limiting mass for the star to survive. If this was the case, it may lead to gravitational collapse of the star (see \cite{kouvaris, zurek}). Therefore, for fermionic DM, we expect $N_{\chi}(t)< N_{\rm Ch}$, where $N_{\rm Ch}\sim (\,M_{\rm Pl}/m_{\chi})^3\sim 1.8 \times 10^{54}\,(\rm1 \,TeV/m_{\chi})^3$ with $M_{\rm Pl}$ the Planck mass, and for the bosonic case $N_{\rm Ch}\sim (\rm M_{\rm Pl}/m_{\chi})^2\sim 1.5 \times 10^{32}\,(\rm 1 \,TeV/m_{\chi})^2$. In case a Bose-Einstein condensate is considered \cite{bramante} $N_{\rm BEC}\simeq 10^{36}\,(T/10^5\, \rm K)^5$  and the condition is $N_{\chi}(t) < N_{\rm Ch}+N_{\rm BEC}$. As described, in the fermionic case, DM remains at all times below the limiting mass, but this may not be the case in the cooling path of the PNS if a Bose-Einstein condensate is formed for a DM particle in the $\sim$TeV mass range. The scenario described here may be indeed at the border of the collapse case, however we will restrict our discussion  to the precollapsed state, leaving the possibility of additional complexity for further investigation.

We can estimate the number of DM decays in the NS phase within a time interval $\Delta t =t-t_{\rm col}\ll\Gamma^{-1}$ using a linear approximation in the exponential expansion in Eq.(\ref{eqnchi}) and with aid of Eq.(\ref{f}),
%
$N_{D,\chi}=N_{\chi} (t_{\rm col})f_{\chi}{\Gamma}\Delta t.
$
%
For a time interval comparable to known ages of ancient pulsars (like that estimated for the isolated pulsar PSR J0108-1431 $\Delta t \simeq \tau_{\rm old\,NS}=1.7\times 10^{8}$ yr \cite{damico}), decays may have profound implications. 

Neglecting any phase space blocking effects, the number of decays assuming decay times similar to those in the cosmic positron/electron anomaly $\tau_{e^+ e^-}$, is given in the present scenario by $N_{D,\chi}=4.2\times { 10^{26}}\left(\frac {f_{\chi}}{2\times 10^{-3}}\right)\left(\frac{1\, \rm TeV}{m_{\chi}}\right)\left(\frac{10^{26}\, s}{\tau_{e^+ e^-}}\right)\left(\frac{\Delta t}{\tau_{\rm old\,NS}}\right)$. This number of decays over the NS lifetime may pose a problem if there is sufficient energy deposited in the nuclear medium to trigger further microscopic effects that may affect the stellar stability, as we will argue. Thus, the possible implications of the absence of these observations may indeed serve to constrain the nature of such decays.

If we now focus on the typical decay final states of  interest for fermionic or bosonic (neutral) DM, we can estimate the energy deposition in the medium. Strictly speaking, injection and deposition are related by an injection fraction that remains unknown since we do not know the preferred decay channels. In this work and in order to compute the size of  the effect, we will consider the photon contribution to decays by two-body channels with intermediate (massive) state daughter particles, quarks, leptons, weak bosons ${\Phi}_w$ or, more generic ${\Phi}$ bosons and photons. Reactions include $\chi \rightarrow {\Phi}_w {\Phi}_w,  l^+l^- , q^+q^-, 2\Phi, \Phi \gamma$, $\chi \rightarrow \Phi_w l$, keeping in mind that more generic decay final states \cite{bell} may well happen. Using the photon spectrum $\frac{dN_{\gamma}}{dE}$ from \cite{pppc4,fortin}, we estimate the injection rate per unit volume and unit energy from the contribution of each channel with corresponding decay rate $\Gamma_i$ at stellar radial location $r$ as $Q(E, r)=n_{\chi}(r) \sum_i \Gamma_i \frac{dN^i_{\gamma}}{dE}$. Then the energy rate injected in the prompt decay channels is written as
\begin{equation}
\frac{dE}{dt}=\int \int E Q(E, r) dE dV.
\end{equation}
Energy release from DM decay is injected locally as microscopic sparks \cite{perez10} in the inner NS core over a central volume $V_{\rm th}=\frac{4}{3}\pi r_{\rm th}^ 3$, where heating and cooling processes compete. As a result, in the thermal volume the average energy density in a time interval $\Delta t$ is
\begin{equation}
\langle{u_{\rm decay}}\rangle \simeq \Delta t \int E Q(E, r) dE.
\end{equation}

However, for single events the energy deposit in a tiny local volume $\delta V$ can be much larger ${u_{\rm decay}}\simeq E^i_{\rm spark}/\delta V$ for each channel and $E^i_{\rm spark}\simeq \int E \frac{dN^i_{\gamma}}{dE}dE$. Indeed DM decay may be regarded as a spark-seeding mechanism in similar fashion to modern versions of direct detection nucleation experiments (PICO, MOSCAB) based on Seitz's theory on {\it heat spike} triggering \cite{seitz}. Thermally induced  quark bubble nucleation in the context of the appearance of quark stars has been already suggested \cite{horvath} and some studies \cite{bubble1} conclude that quark matter bubbles may nucleate if the temperature locally exceeds $\delta T\simeq$ few MeV, provided the MIT model bag constant is $B^{1/4}=150\pm 5$ MeV.

 In the scenario depicted here, decays may provide the energy injection to create a bubble. In order to see this, we estimate the minimum critical work \cite{bubble1} needed to nucleate a neutral stable spherical quark bubble in the core of the cold NS. It is given by $W_c=\frac{16 \pi}{3} \sqrt{\frac{2 \gamma^{3}}{\Delta P}}$ where $\Delta P=P_q-P_n$ and $P_q$ ($P_n$) is the quark (nucleon) pressure. For a two-flavour ud-quark system  $P_{q}=\sum\limits_{i=u,d} \frac{\mu^4_i}{4\pi^2}-B$ and assuming a neutron-rich system $P_n\simeq \frac{(\mu^2_n-m^2_n)^{5/2}}{15\pi^2m_n}$ as most of the pressure will effectively be provided by neutrons in the n fluid.  $\gamma=\sum\limits_{i=u,d} \frac{\mu^2_i}{8\pi^2}$, is the curvature coefficient and $\mu_{i}$ $(\mu_n$) is the quark (nucleon) chemical potential related to the Fermi momentum of the degenerate system $\mu_i=p_{F\,i}$ ($\mu_n=\sqrt{m^2_n+p^2_{F\,n}}$). Electrical charge neutrality requires for the ud matter $n_d=2 n_u$ and $n_n=(n_u+n_d)/{3}$ with $n_i= \frac{\mu^3_i}{\pi^2}$ in the light quark massless limit. Note that we do not include further refinements due to non-zero quark masses, in-medium or Coulomb effects or droplet surface tension since they do not change the global picture as we want to keep a compact meaningful description of the nucleation process.
Bubbles have at least a radius $R_c=\sqrt{2 \gamma/\Delta P}$ and their stability is granted as they reach the minimum baryonic number $A_{\rm min}\sim 10$ \cite{amin} when $A\sim \frac{4}{3} \pi R^3_c n_n > A_{\rm min}$.
Although there may be additional efficiency quenching factors we expect that they do not  significantly alter the picture presented here but we discuss this possibility below as a possible source of uncertainty.

In previous work by Harko et al. \cite{bubble2}, it was assumed that to convert the full NS, at least one stable capable-to-grow quark bubble should be formed. However, in order to be conservative we will require a number $N_0$ of bubbles that could lead to a sizeble amount of nucleated matter, then one can write this condition using the spark seeding rate as
\begin{equation}
{N}_{\rm bub} \simeq \int \frac{dN_{\rm bub}}{dE} \frac{dE}{dt} dt \geq N_0.
\end{equation} 
 If this was indeed the scenario, a possibly catastrophic NS to quark star transition could occur when the macroscopic deconfinement proceeds via detonation modes to rapidly consume the star  \cite{nieb}. The GRB signal emitted has been estimated in \cite{perez13} and subsequent emission in  cosmic ray channels is also expected \cite{kumiko-perez-silk}.

In our galaxy, the supernova rate is about $R=10^{-2}\, \rm yr^{-1}$ so that an average rate of NS  formation over the age of the universe $\tau_U\sim 4.34 \times 10^{17}$ s yields $N_{NS}\sim R \tau_U\sim 10^{8-9}$. Assuming this population is formed by regular nucleonic NSs, and under the conditions discussed so far regarding DM properties a lower limit optimistic value (provided one bubble is formed, $N_0=1$) for $\tau_{\chi}$ can be accordingly set from the age of the old NSs as  
$\tau_{\chi}\gtrsim N_{\chi} (t_{\rm col})f_{\chi}\Delta t_{\rm old\,NS}$.

The energy density necessary to create such a quark bubble with volume $V_d\simeq \frac{4}{3} \pi R^3_c$ is therefore $u_{\rm bub} \simeq W_c/V_d\simeq 5.4\times 10^{35}$ $\rm erg/cm^{3}$. This estimate is in agreement with similar and more detailed calculations \cite{bubble2}. To allow the quark bubbles to  nucleate,  the local energy densities must then fulfill $u_{\rm decay} \gtrsim u_{\rm bub}$. Some attemps to  computationally model progression of seeds of quark matter inside NSs have been recently performed in \cite{herzog}.

\begin{figure}
\begin{minipage}[b]{1.\linewidth} 
\centering
\includegraphics[scale=0.95]{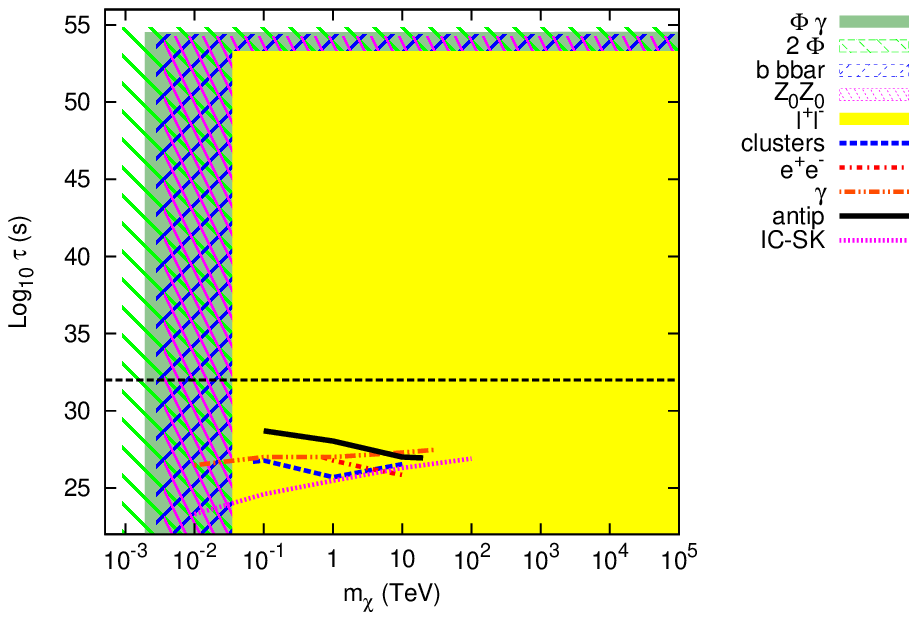}
\end{minipage}
\hspace{0.25cm} 
\begin{minipage}[b]{1.\linewidth}
\centering
\includegraphics[scale=0.95]{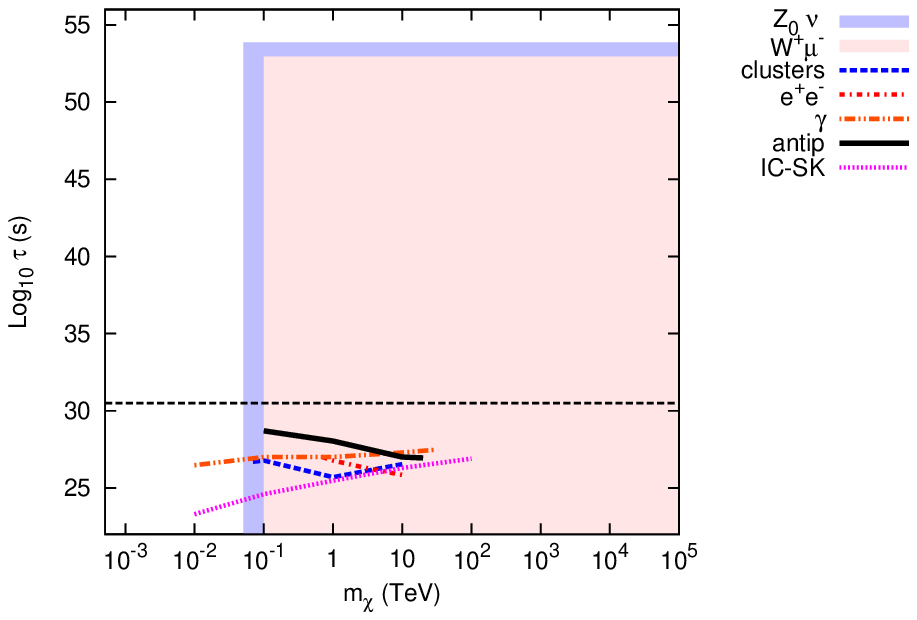}
\caption{Logarithm of DM decay time as a function of mass. Upper (lower) panel depicts bosonic (fermonic)  channels. Colored regions show our excluded values assuming $N_0=1$. We show current constraints from upper limits coming from gamma rays in the Galactic Center and in galaxy clusters, positron and antiproton fluxes, and IceCube and Super Kamiokande. Dashed lines on both panels denote our most pessimistic global lower bound on decay time arising from the uncertainties in the percolation scenario (see details in the text).} 
\label{Fig1}
\end{minipage}
\end{figure}

In Fig.\ref{Fig1} we plot the logarithm of $\tau_{\chi}$ versus $m_{\chi}$ for bosonic (upper panel) and fermionic (lower panel) DM particle cases. We consider the optimistic value $N_0=1$ where one single stable bubble can proceed to convert the full star \cite{bubble2}. The different partially overlapping rectangular-shaped colored regions (solid or hatched) represent exclusion regions for each of the decay channels considered in this work. Particle decays in this region  would produce NS transitions over ages below those assumed for regular old NSs. We assume a DM density $\rho^{ambient}_{\chi,0}\simeq 0.3$ $\frac{\rm GeV}{cm^3}$ and, in line with our previous discussion, we set $n_n=5n_0$. 
We can see there is a threshold mass below which energy injection is not able to grow stable bubbles.  We find that for masses $(m_{\chi}/ \rm TeV) \gtrsim 9 \times 10^{-4}$  or  $(m_{\chi}/ \rm TeV) \gtrsim 5 \times 10^{-2}$ in the  bosonic or fermionic cases respectively, lifetimes ${\tau_{\chi}}\lesssim 10^{55}$ s or ${\tau_{\chi}}\lesssim 10^{53}$ s accordingly, are excluded. 
As already mentioned, there is a natural limit for NSs to effectively test decaying DM, from their ability to accrete and retain DM during its lifetime $\tau_{\rm old\,NS}$. We can use this result to set exclusion regions for $\tau_{\chi}$  complementary to those shown in other works under the conditions of validity of our scenario. We plot upper limit constraints coming from  $e^+e^-$, gammas, antiproton fluxes and IceCube and SuperKamiokande \cite{ibarra13, dugger}. Note that bosonic decay channels are able to probe a wider range of DM masses as compared to fermionic decays. Our optimistic limits are considerably stronger than earlier results, including the CMB constraints \cite{diamanti}. 

At this point we should critically asses the several uncertainties that may affect  our estimate. As a result we depict a global lower bound on decay time in dashed lines in Fig.\ref{Fig1}. In the following we explain the different origin of the proposed reduction factors in the DM lifetimes. 

In the mechanism of bubble nucleation inside the core of NSs presented in this work, a  major uncertainty concerns quenching efficiencies, which may  require more than a single bubble to trigger the NS collapse. In a more conservative scenario,  the minimum number of bubbles to effectively distort the internal NS structure may be linked to the mechanical instability one can create by nucleation. The associated variation in pressure is proportional to the volume affected or number of bubbles, this is estimated as given by $\delta P\simeq \left[\frac{\partial P}{\partial N_{\rm bub}}\right]_0 \delta N_{\rm bub}$ where the coefficient is related to regular nuclear matter compressibility (in the absence of bubbles). In order to quantify the effect, let us consider an old and cold NS with $T\sim 10^5$ K and $\rho_n= 5 \rho_0$ for a DM particle with $m_{\chi}\sim1$ TeV.
The DM particle number inside the NS thermal volume is ${A_{{\rm th}}}\simeq V_{\rm th} n_n \simeq 10^{42}$. If each possibly formed bubble has a baryonic number $A_{\rm min}\simeq 10$ then a macroscopic $N_0$ number of bubbles acount for $A_{\rm bub}\sim 10 N_0$. In principle, a standard variation in the number of nucleons depleted from the thermal volume nucleon sea, $\sqrt{A_{{\rm th}}}\simeq 10^{21}$, could provide a large enough variation in pressure to trigger transitioning. Then the number of bubbles to create this perturbation is $N_0\simeq \sqrt{A_{{\rm th}}}/A_{\rm min}\simeq 10^{20}$.  In such a pessimistic scenario, the limits on DM decay lifetimes would be reduced  by a similar proportion. 

 In addition, heavy-ion collision simulations using perturbative QCD \cite{eskola} give estimates of the typical quenching factor or ratio of energy spread in this context as $Q\simeq \cal{O}$(0.1). One must note however that the jet-sized regions in heavy-ion collisions are about three orders of magnitude or more  smaller in local energy density i.e. $\sim$ $\rm GeV/fm^3$ than the ranges presented in this work with $\sim$ a few fm spatial spread on a time-scale of several fm/c. This size is comparable to typical bubble sizes. We expect that this correction does not significantly affect the results presented here. In addition one should consider that Pauli blocking effects of ordinary matter may reduce the kinematical phase space available for the DM-n interaction effects at the few percent for low $m_{\chi}$ while being negligible in the TeV range and above. Also the uncertainty on the quoted minimum bubble size and baryonic number can somewhat further constrain the lower $m_{\chi}$ range, and  so we should be safe for somewhat larger masses. Note however that some works have suggested a decaying PeV mass DM particle as an explanation to IceCube PeV cascade data \cite{esmaili,feldstein}. Such a  PeV mass DM particle would be however at the limit of thermalization for $\sigma_{\chi n}\sim 10^{-44} \,\rm cm^2$, a smaller cross-section invalidating the thermalization procedure quoted in our NS scenario and cannot be ruled out in light of our current findings. However, existing data, as depicted in Fig.\ref{Fig1} allows for a $\sim 10^4-10^5$ factor in lifetime to vary proportionally the cross-section and DM mass parameter space, and self-consistently other parameters in the model. As we do not intend to perform a detailed analysis in this work, we leave this task for future contributions.
Linked to this argument, we have considered an average-sized massive NS, however, variations in mass and radius in current surveys are of the order of $30\%$ \cite{Kiziltan} and the capture rate would be affected in a similar fashion, resulting in a moderately reduced capture efficiency.

In  summary, we have shown that the current population of galactic NS  may constrain the nature of a decaying bosonic or fermionic DM  particle with mass in the GeV-TeV range. Since there is a number of uncertainties intrinsic to the model, we discuss in particular two cases, namely the optimistic case where we find that DM particles with lifetimes ${\tau_{\chi}}\lesssim 10^{55}$ s exclude masses $(m_{\chi}/ \rm TeV) \gtrsim 9 \times 10^{-4}$  or ${\tau_{\chi}}\lesssim 10^{53}$ s excludes  $(m_{\chi}/ \rm TeV) \gtrsim 5 \times 10^{-2}$ in the  bosonic or fermionic cases, respectively. Uncertainties associated may substantially reduce these limits. In the pessimistic case, a global bound is set with lifetimes a factor  $\gtrsim 10^{20}$ smaller that still improve on rival bounds from $e^+e^-$, $\gamma$, $\bar p$ fluxes and IceCube and SuperKamiokande, leaving a combined factor of $\sim 10^4-10^5$ in lifetime if different cross-sections or masses are considered. These results are obtained from the prior of avoiding nucleation of quark bubbles in a NS core due to efficient energy injection by spark seeding. If this was the case, a conversion from NS into quark star would be triggered, thereby reducing the population of  regular NS in the galaxy.

M.A.P.G. would like to thank useful discussions with R. Lineros, C. Mu\~noz and P. Panci. She acknowledges the kind hospitality of IAP where part of this work was developed. This research has been supported at University of Salamanca by the Consolider MULTIDARK and FIS2012-30926 MINECO projects and at IAP by  the ERC project  267117 (DARK) hosted by Universit\'e Pierre et Marie Curie - Paris 6   and at JHU by NSF grant
OIA-1124403.

\end{document}